\documentclass[useAMS,usenatbib]{mnras}
\usepackage{array}
\usepackage{times}
\usepackage{graphicx}
\usepackage{mathptmx}
\usepackage{amsmath}
\usepackage{amssymb}
\usepackage[usenames]{color}
\def\mnras{MNRAS}
\def\apjl{ApJL }
\def\aj{AJ }
\def\apj{ApJ }

\def\apjs{ApJS }
\def\araa{ARAA }
\def\aap{A\&A }
\def\nat{Nature }
\usepackage{hyperref}
\hypersetup{breaklinks=true,colorlinks=true,citecolor=blue,linkcolor=purple,filecolor=black,runcolor=black}

\usepackage{etoolbox}
\makeatletter
\patchcmd\@combinedblfloats{\box\@outputbox}{\unvbox\@outputbox}{}{%
   \errmessage{\noexpand\@combinedblfloats could not be patched}%
}%
 \makeatother
 
\begin{document}
\newcommand{\comment}[1]{}
\definecolor{purple}{RGB}{160,32,240}
\newcommand{\peter}[1]{\textcolor{purple}{(\bf #1)}}
\newcommand{\macc}{M_\mathrm{acc}}
\newcommand{\mpeak}{M_\mathrm{peak}}
\newcommand{\mnow}{M_\mathrm{now}}
\newcommand{\vacc}{v_\mathrm{acc}} 
\newcommand{\vpeak}{v_\mathrm{peak}} 
\newcommand{\vnow}{v^\mathrm{now}_\mathrm{max}}

\newcommand{\Mnfw}{M_\mathrm{NFW}}
\newcommand{\Msun}{\;\mathrm{M}_{\odot}}
\newcommand{\mvir}{M_\mathrm{vir}}
\newcommand{\rvir}{R_\mathrm{vir}}
\newcommand{\vmax}{v_\mathrm{max}}
\newcommand{\vmac}{v_\mathrm{max}^\mathrm{acc}}
\newcommand{\mvac}{M_\mathrm{vir}^\mathrm{acc}}
\newcommand{\sfr}{\mathrm{SFR}}
\newcommand{\plotgrace}[1]{\includegraphics[width=\columnwidth,type=pdf,ext=.pdf,read=.pdf]{#1}}
\newcommand{\plotssgrace}[1]{\includegraphics[width=0.95\columnwidth,type=pdf,ext=.pdf,read=.pdf]{#1}}
\newcommand{\plotgraceflip}[1]{\includegraphics[width=\columnwidth,type=pdf,ext=.pdf,read=.pdf]{#1}}
\newcommand{\plotlargegrace}[1]{\includegraphics[width=2\columnwidth,type=pdf,ext=.pdf,read=.pdf]{#1}}
\newcommand{\plotlargegraceflip}[1]{\includegraphics[width=2\columnwidth,type=pdf,ext=.pdf,read=.pdf]{#1}}
\newcommand{\plotminigrace}[1]{\includegraphics[width=0.5\columnwidth,type=pdf,ext=.pdf,read=.pdf]{#1}}
\newcommand{\plotmicrograce}[1]{\includegraphics[width=0.25\columnwidth,type=pdf,ext=.pdf,read=.pdf]{#1}}
\newcommand{\plotsmallgrace}[1]{\includegraphics[width=0.66\columnwidth,type=pdf,ext=.pdf,read=.pdf]{#1}}
\newcommand{\plotappsmallgrace}[1]{\includegraphics[width=0.33\columnwidth,type=pdf,ext=.pdf,read=.pdf]{#1}}

\newcommand{\vmp}{v_\mathrm{Mpeak}}
\newcommand{\fq}{f_\mathrm{Q}}
\newcommand{\sfrq}{SFR_\mathrm{Q}}
\newcommand{\sfrsf}{SFR_\mathrm{SF}}
\newcommand{\sigq}{\sigma_\mathrm{Q}}
\newcommand{\sigsf}{\sigma_\mathrm{SF}}

\newcommand{\hinv}{h^{-1}}
\newcommand{\mpc}{\rm{Mpc}}
\newcommand{\hmpc}{$\hinv\mpc$}

\newcommand{\mstar}{M_\ast}

\title{The Most Massive Galaxies and Black Holes Allowed by $\Lambda$CDM}

\author[Behroozi \& Silk]{Peter~Behroozi$^{1}$,\thanks{E-mail:     behroozi@email.arizona.edu} Joseph Silk$^{2,3}$ 
\\
\\
$^{1}$ Department of Astronomy and Steward Observatory, University of Arizona, Tucson, AZ 85721, USA\\
$^{2}$ Institut d'Astrophysique, UMR 7095 CNRS, Universit\'e Pierre et Marie Curie, 75014 Paris, France\\
$^{3}$ Department of Physics and Astronomy, Johns Hopkins University, Baltimore, MD 21218, USA\\
}
\date{Released \today}
\pubyear{2018}

\maketitle

\begin{abstract}
Given a galaxy's stellar mass, its host halo mass has a lower limit from the cosmic baryon fraction and known baryonic physics.  At $z>4$, galaxy stellar mass functions place lower limits on halo number densities that approach expected $\Lambda$CDM halo mass functions.  High-redshift galaxy stellar mass functions can thus place interesting limits on number densities of massive haloes, which are otherwise very difficult to measure.  Although halo mass functions at $z<8$ are consistent with observed galaxy stellar masses if galaxy baryonic conversion efficiencies increase with redshift, \textit{JWST} and \textit{WFIRST} will more than double the redshift range over which useful constraints are available.  We calculate maximum galaxy stellar masses as a function of redshift given expected halo number densities from $\Lambda$CDM.  We apply similar arguments to black holes.  If their virial mass estimates are accurate, number density constraints alone suggest that the quasars SDSS J1044$-$0125 and SDSS J010013.02$+$280225.8 likely have black hole mass --- stellar mass ratios higher than the median $z=0$ relation, confirming the expectation from Lauer bias.  Finally, we present a public code to evaluate the probability of an apparently $\Lambda$CDM-inconsistent high-mass halo being detected given the combined effects of multiple surveys and observational errors.
\end{abstract}

\section{Introduction}

\label{s:introduction}

In the framework of Lambda Cold Dark Matter ($\Lambda$CDM), galaxies form at the centres of dark matter haloes \citep[see][for reviews]{Silk12,Somerville15}.  The ratio of galaxy stellar mass to halo mass has an absolute maximum at the cosmic baryon fraction ($f_b \sim 0.16$; \citealt{Planck15}).  In practice, stellar feedback processes limit the maximum fraction of baryons converted to $\lesssim 40\%$ \citep{moster-09,Moster12,Moster18,Behroozi10,BWC13} even when adopting a \cite{Salpeter55} initial mass function (IMF).  At $z<4$, this maximum fraction is never achieved for massive haloes ($M_h>10^{12}\Msun$), due to inefficient cooling \citep{Lu11b} and feedback from supermassive black holes \citep{Silk98}.
  At $z>4$, however, comparisons of galaxy and halo number densities suggest that massive haloes can reach from 10--40\% typical integrated efficiencies in converting baryons into stars, again depending on assumptions for the IMF and luminosity---stellar mass conversions \citep{BWC13,BehrooziHighZ,Finkelstein15,Sun16,Moster18}.

Conversely, an observed galaxy mass ($M_\star$) places a lower limit on its host halo mass ($M_h$).  $\Lambda$CDM alone implies that $M_h > M_\star / f_b \sim 6.3 M_\star$, and known baryonic physics would give more stringent limits depending on the assumed maximum conversion efficiency.  This fact has been used in \cite{Steinhardt16} to argue that galaxy number densities at $z\sim5-6$ are already inconsistent with $\Lambda$CDM.  Although we disagree with their assumptions  (especially that the $M_\star / M_h$ ratio cannot increase at $z>4$) and therefore also their conclusions, the basic principle that galaxy number densities constrain halo number densities is well-established.

As galaxy number densities are consistent with halo number densities for redshifts $z\lesssim 8$ \citep{BWC13}, we compute galaxy mass limits corresponding to expectations from 
typical $\Lambda$CDM baryon fraction limits over $7<z<20$, observable with future infrared space-based telescopes (e.g., \textit{JWST}, the \textit{James Webb Space Telescope}, and \textit{WFIRST}, the \textit{Wide-Field InfraRed Survey Telescope}).

Similarly, useful physical thresholds can be calculated for supermassive black holes (SMBHs).  The number density of a given quasar sample places a lower limit on the number density of their host haloes, which in turn limits the maximum average host halo mass in $\Lambda$CDM \citep[see also][]{Haiman01}.  This then limits the maximum average host galaxy mass (via $M_\star < f_b M_h$).  Hence, given the number density of black holes above a certain mass, we can derive a lower limit for their $M_\bullet / M_\star$ ratios without requiring observations of the host galaxy.  The current claimed maximum $M_\bullet / M_\star$ ratio is 15\% \citep{vdBosch12,Seth14}.  In comparison, the highest median relations for the $M_\bullet / M_\star$ ratio at $z=0$ \citep{Kormendy13,Savorgnan16} give $\sim 0.4$\% for $M_\mathrm{bulge} = 10^{11}\Msun$ (after conversion to a \citealt{Salpeter55} IMF).  

Throughout, we assume a flat, $\Lambda$CDM cosmology with $\Omega_M = 0.309$, $\Omega_b = 0.0486$, $\sigma_8=0.816$, $h=0.678$, $n_s = 0.967$, corresponding to the best-fit \textit{Planck} cosmology \citep{Planck15}, as well as a \cite{Salpeter55} IMF.  For halo masses, we use the virial overdensity definition of \cite{mvir_conv}.

\begin{figure}
\vspace{-5ex}
\plotgrace{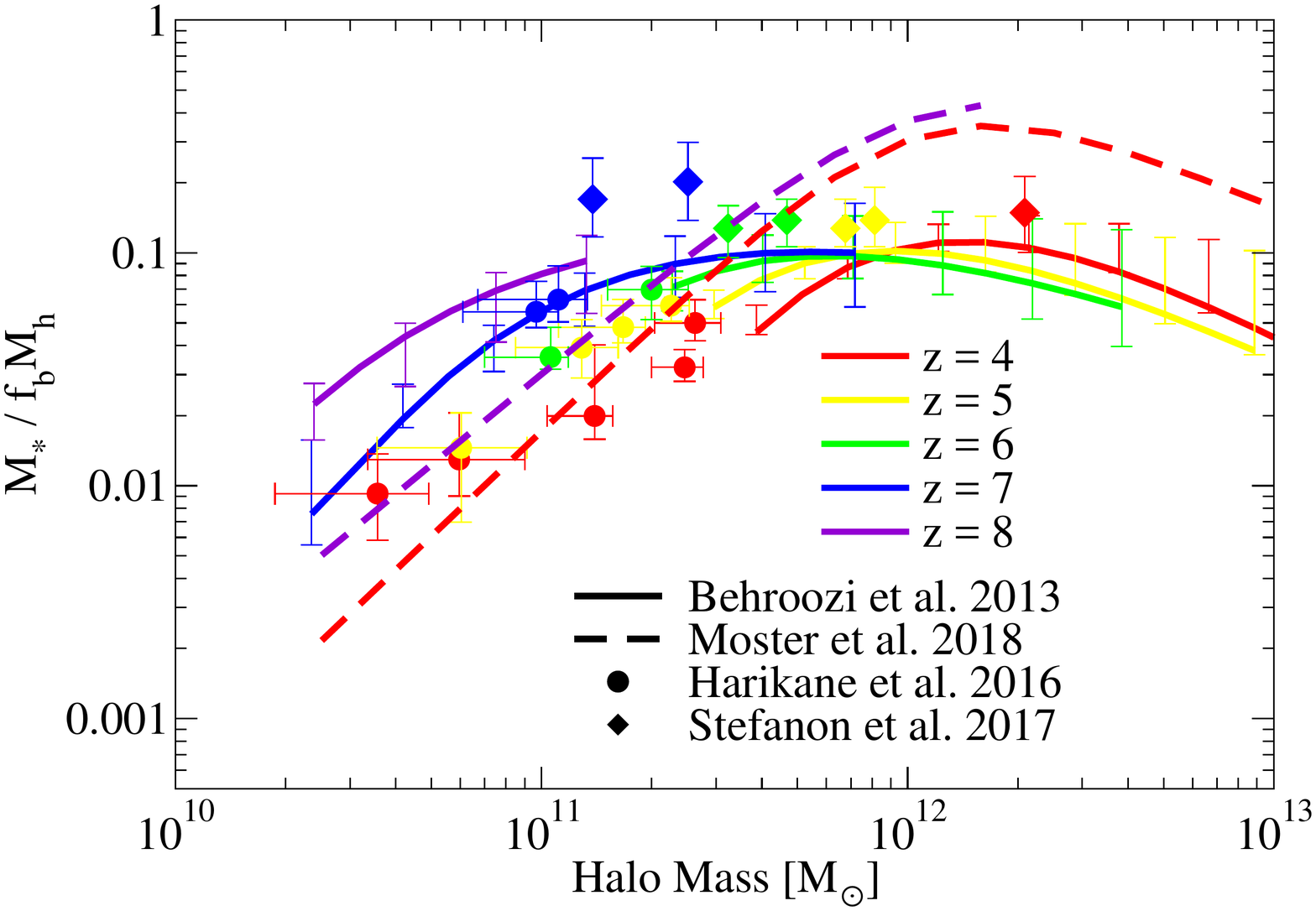}\\[-5ex]
\caption{Median stellar mass--baryonic mass ratios at $z=4-8$ reach up to 10-40\%.  With a scatter of even 0.2 dex (as at low redshifts), it is plausible that individual galaxies can reach ratios near unity.  Results have been converted to a \protect\cite{Salpeter55} IMF and to Planck cosmology where appropriate.  \protect\cite{BWC13}, \protect\cite{Stefanon17}, and \protect\cite{Moster18} use abundance matching; \protect\cite{Harikane16} uses halo occupation distribution modeling of angular correlation functions.}
\label{f:smhm}
\end{figure}

\begin{figure*}
\vspace{-6ex}
\plotgrace{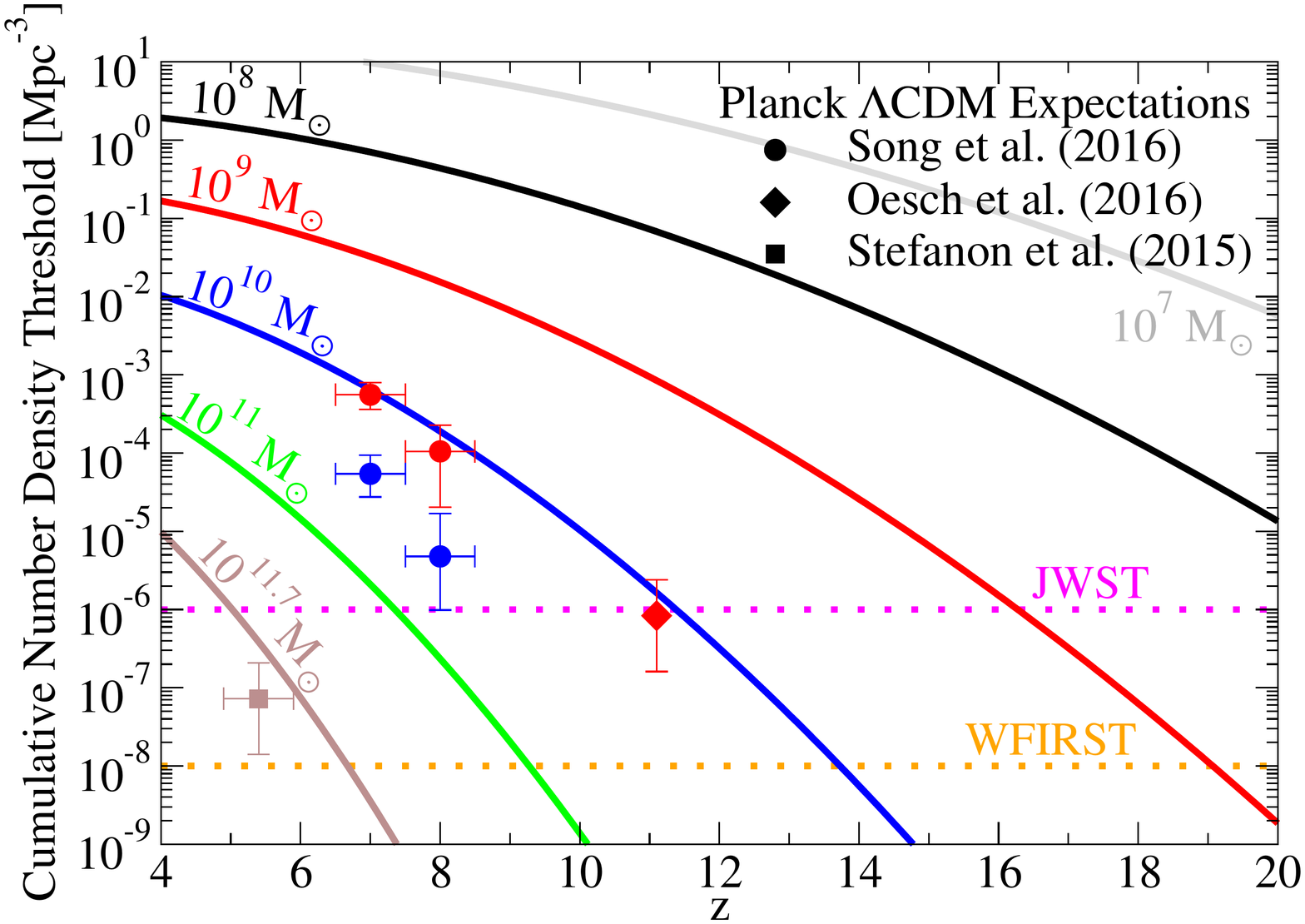}
\plotgrace{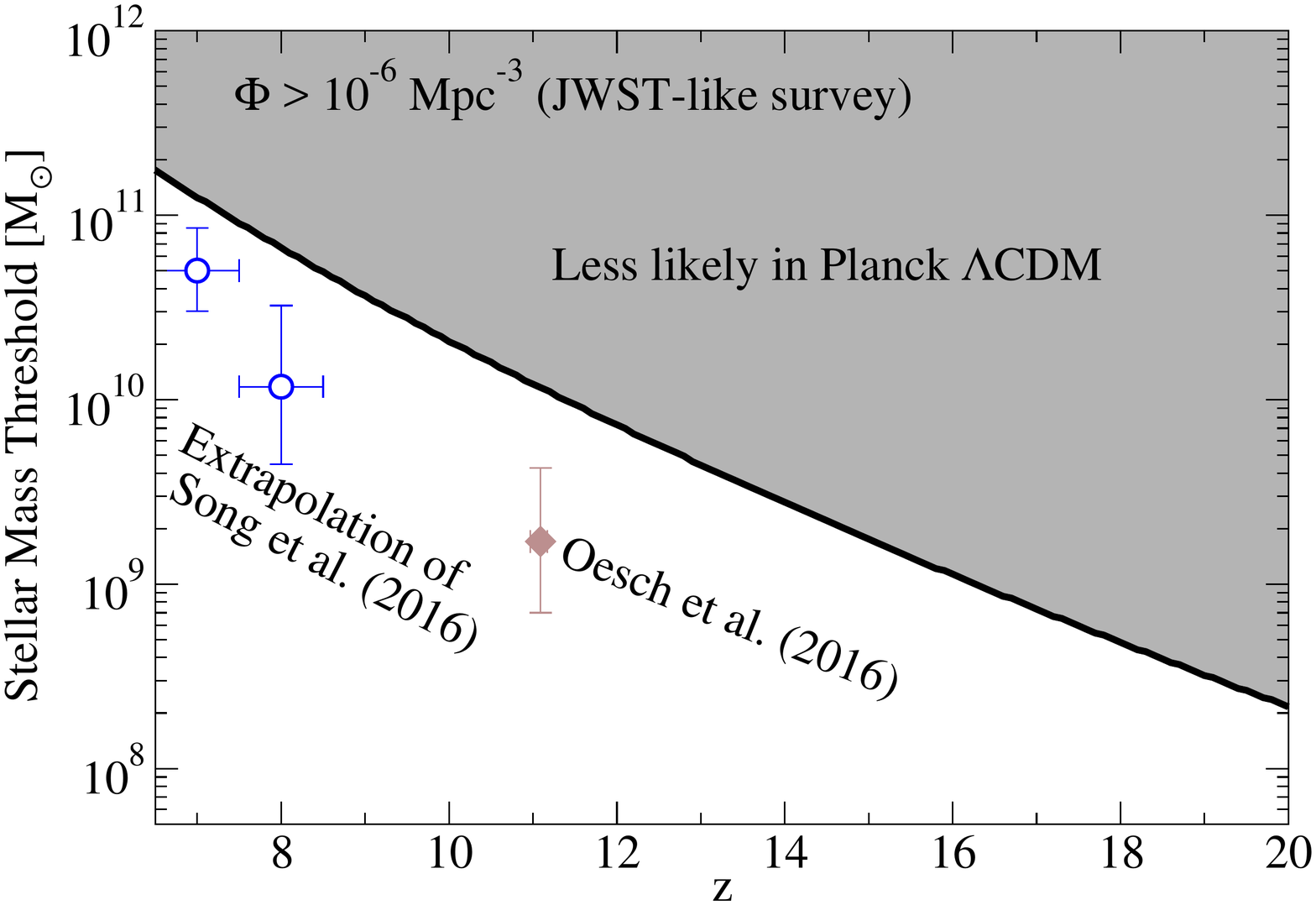}\\[-7ex]
\plotgrace{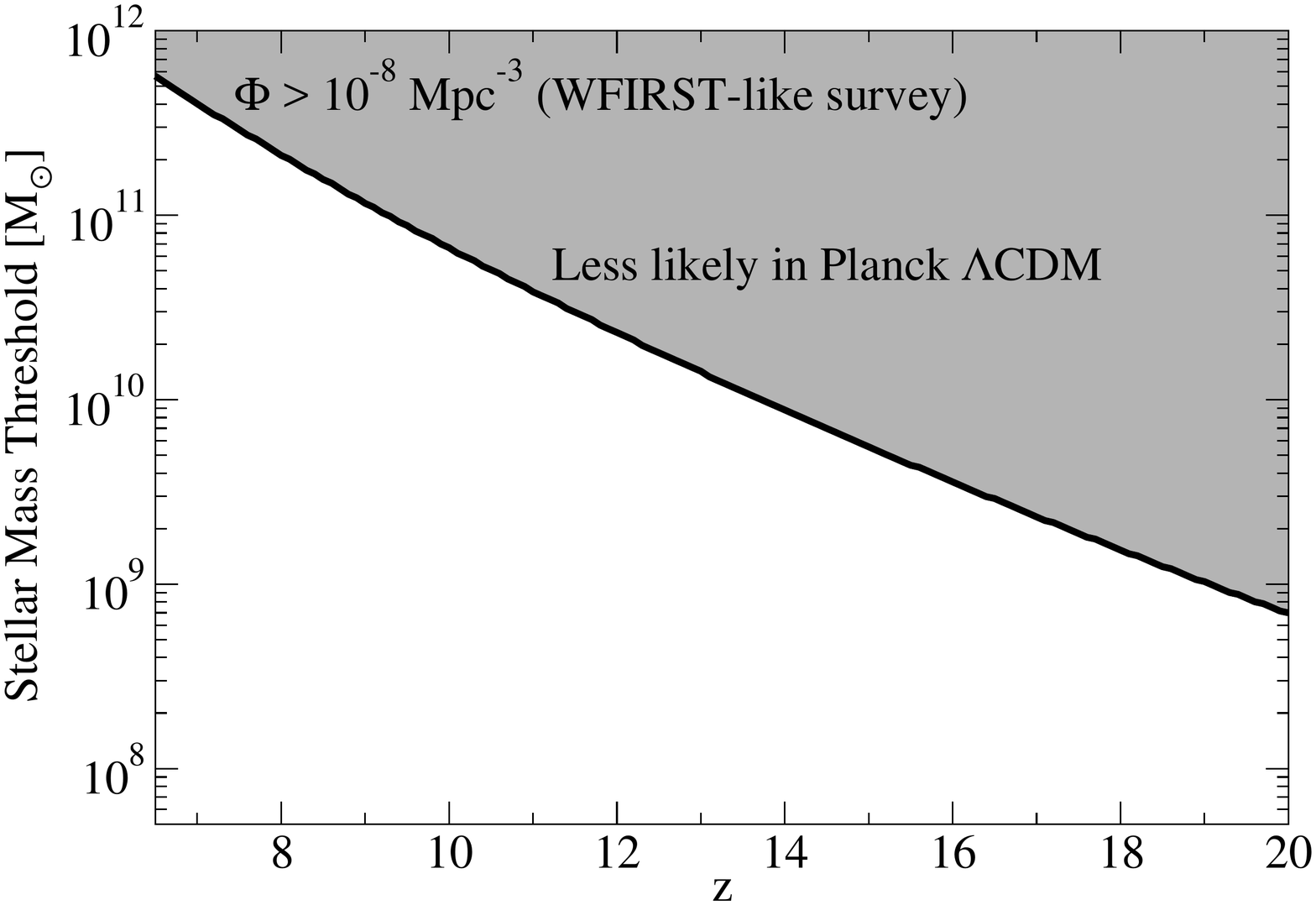}
\plotgrace{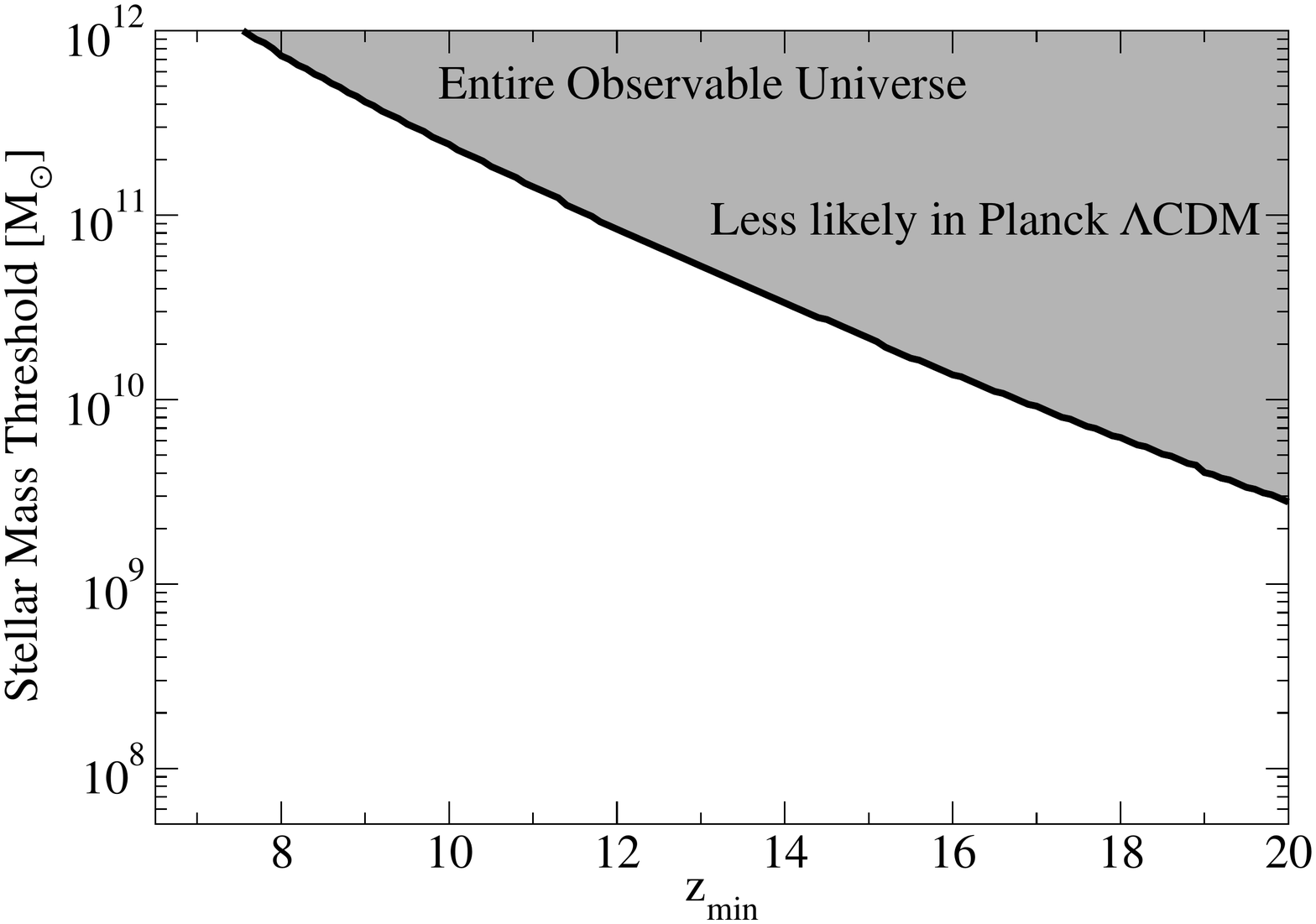}\\[-5ex]
\caption{\textbf{Top-left} panel: Cumulative number density thresholds as a function of stellar mass and redshift; observed galaxy cumulative number densities are expected to be below these thresholds in Planck $\Lambda$CDM, subject to sample variance and observational errors (see Appendix \ref{a:code}).  Colored solid lines correspond to different stellar mass thresholds; the \textit{brown} line corresponds to the $10^{11.7}\Msun$ mass estimated for the massive galaxy in \protect\cite{Stefanon15}.  Colored dotted lines correspond to expected number density limits for the \textit{JWST} and \textit{WFIRST} missions.  \textbf{Top-right} panel: threshold stellar masses for a cumulative number density of $\Phi = 10^{-6}$ Mpc$^{-3}$.  If a survey found that galaxies with stellar masses larger than the \textit{black} line had a cumulative number density higher than $10^{-6}$ Mpc$^{-3}$ with significant confidence (see Appendix \ref{a:code}), it would rule out $\Lambda$CDM.   \textbf{Bottom-left} panel: same, for a cumulative number density threshold of $\Phi = 10^{-8}$ Mpc$^{-3}$.  \textbf{Bottom-right} panel: same, for the entire observable Universe (i.e., all sky survey with $z_\mathrm{min} < z < \infty$).}
\vspace{-5ex}
\label{f:mass}
\end{figure*}

\begin{figure*}
\vspace{-3ex}
\plotgrace{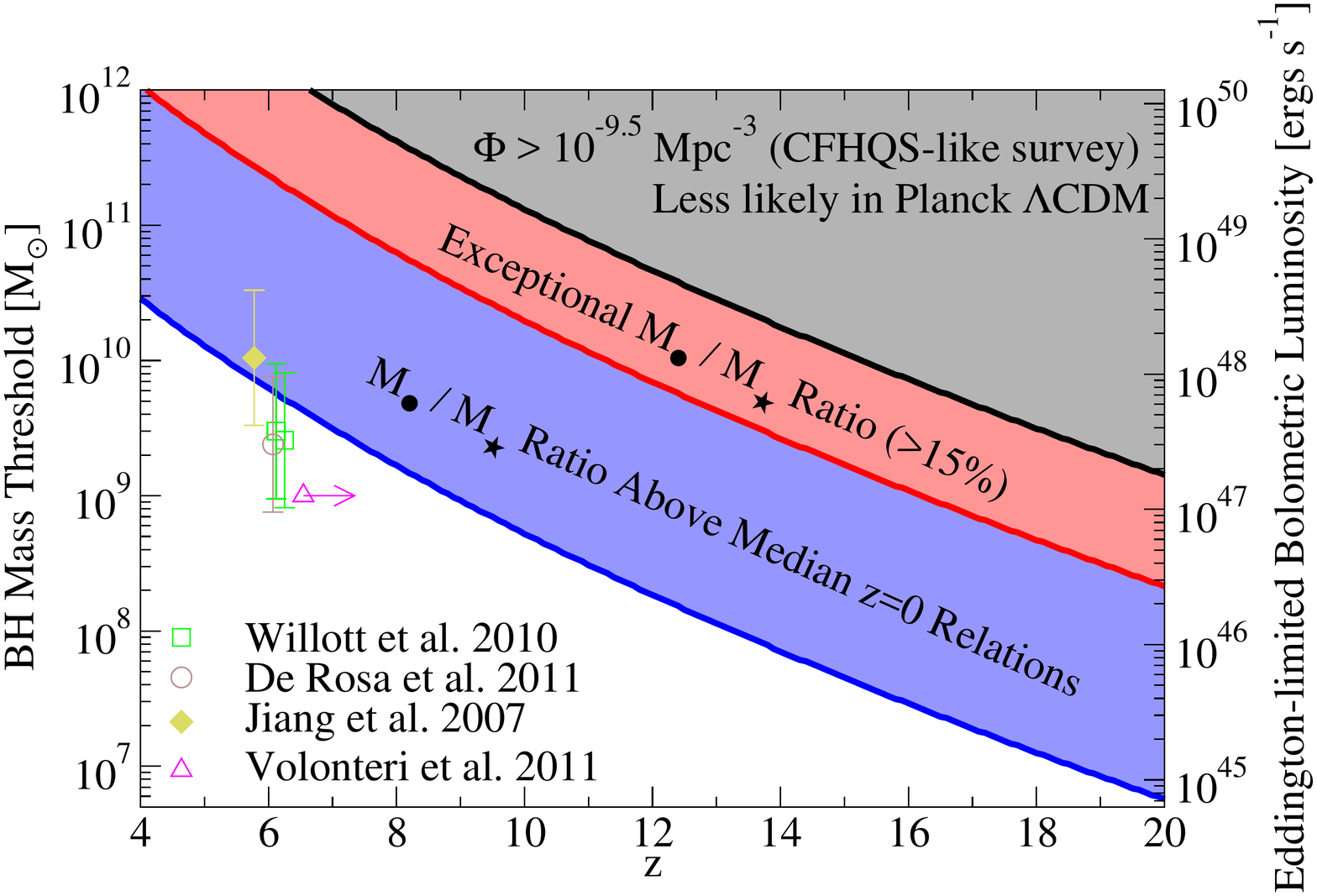}
\plotgrace{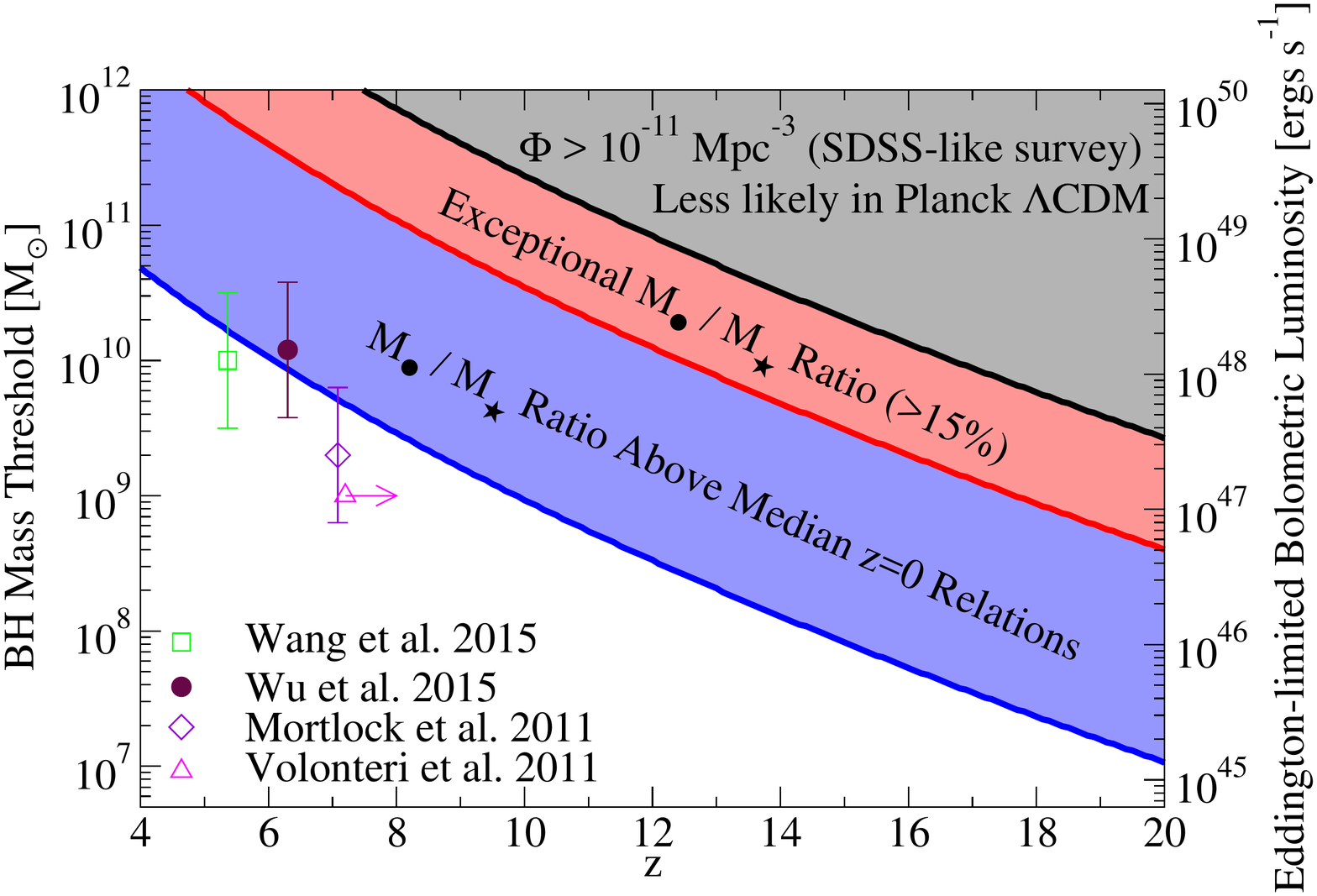}\\[-5ex]
\caption{\textbf{Left} panel: threshold black hole masses for a cumulative number density of $\Phi = 10^{-9.5}$ Mpc$^{-3}$.  If a survey found black holes with masses above the \textit{red} line and cumulative number densities above $10^{-9.5}$ Mpc$^{-3}$, those black holes would exceed the current record $z=0$ black hole mass---stellar mass ratio.  Similarly, if a survey found black holes with masses above the \textit{blue} line and cumulative number densities above $10^{-9.5}$ Mpc$^{-3}$, those black holes would exceed all current determinations of the median $z=0$ black hole mass---stellar mass ratio. \textbf{Right} panel: same, for a cumulative number density threshold of $\Phi = 10^{-11}$ Mpc$^{-3}$.  Solid data points indicate the most massive black holes found to date in surveys of the respective volume; open data points indicate less-massive black holes.}
\label{f:mass_bh}
\end{figure*}

\section{Methodology}

\label{s:methodology}

We adopt cumulative halo mass functions ($\Phi_h$) from \cite{BWC13} and define a redshift-dependent maximum cumulative galaxy mass function:
\begin{eqnarray}
\Phi_\mathrm{\star,\Lambda{}CDM}(M_\star,z) & \equiv & \Phi_h(M_\star / f_b,z)
\end{eqnarray}
$\Phi_\mathrm{\star,\Lambda{}CDM}$ is the $\Lambda$CDM upper limit on the true stellar mass function---equivalently, it is the expected collapsed gas mass function for $\Lambda$CDM.  In practice, a given survey volume may contain a higher number density due to sample variance or observational errors, especially when multiple surveys are conducted.  Appendix \ref{a:code} offers a simple method and code to test whether a given outlier is significantly discrepant from the $\Lambda$CDM prediction.

Given that median stellar mass fractions already reach up to $40\%$ at $z>4$ (Fig.\ \ref{f:smhm}) with the assumption of a \cite{Salpeter55} IMF, and that measurements of scatter in stellar mass at fixed halo mass are typically $\sim0.2$ dex \citep[e.g.,][]{Reddick13}, it is plausible that some galaxies could reach nearly $100\%$ conversion efficiencies modulo the effects of stellar mass loss (20-30\%; \citealt{Conroy09}).

We also define two cumulative supermassive black hole (SMBH) mass functions:
\begin{eqnarray}
\Phi_\mathrm{\bullet,max}(M_\bullet,z) & \equiv & \Phi_h(M_\bullet / (0.15 f_b),z)\\
\Phi_\mathrm{\bullet,median}(M_\bullet,z) & \equiv & \Phi_h(M_\bullet / (0.004 f_b),z)
\end{eqnarray}
If $\Phi_\mathrm{\bullet,obs}$ exceeds $\Phi_\mathrm{\bullet,max}$, the black holes must have observed $M_\bullet / M_\star$ ratios of $>$15\%, regardless of galaxy formation physics.  If $\Phi_\mathrm{\bullet,obs}$ exceeds $\Phi_\mathrm{\bullet,median}$, then some of the SMBHs must have observed $M_\bullet / M_\star$ ratios above the $z=0$ median relation, again regardless of galaxy formation physics.  Specifically, one expects at least a fraction
\begin{equation}
f_\uparrow \equiv  \frac{\Phi_\mathrm{\bullet,obs}-\Phi_\mathrm{\bullet,median}}{\Phi_\mathrm{\bullet,obs}}
\end{equation}
of the SMBHs to have higher-than-median $M_\bullet / M_\star$ ratios.  This is not necessarily a surprising finding; indeed, any $M_\bullet$-selected sample will have at least half of the sample higher than the median relation.  However, a high $f_\uparrow$ could suggest evolution in the $M_\bullet / M_\star$ relation or in the scatter in that relation (including observational errors) to high redshifts.

\section{Results}
\label{s:results}

For arbitrary future surveys, Fig. \ref{f:mass} also shows cumulative number density thresholds as a function of galaxy stellar mass from $z=4$ to $z=20$, including a comparison to the massive galaxies in \cite{Stefanon15} and \cite{Oesch16} and to the stellar mass functions in \cite{Song15}.  For \cite{Oesch16}, we use their estimated search volume of $1.2\times 10^6$ Mpc.  As in Fig.\ \ref{f:smhm}, more massive galaxies tend to reach higher $M_\ast / M_h$ ratios, so their number densities more closely approach the expected collapsed gas mass function from $\Lambda$CDM.

Similar to CANDELS \citep{Koekemoer11,Grogin11} with \textit{Hubble}, a future \textit{JWST} survey may probe galaxy cumulative number densities down to $n_J \sim 10^{-6}$ Mpc$^{-3}$.  \textit{WFIRST} has a $\sim 100\times$ larger field of view, so it may reach cumulative number densities of $n_W \sim 10^{-8}$ Mpc$^{-3}$.  For these two threshold densities, we plot $\Phi^{-1}_\mathrm{\star,\Lambda{}CDM}(n,z)$ (i.e., the stellar mass $M_\star(z)$ at which $\Phi_\mathrm{\star,\Lambda{}CDM}(M_\star(z), z) = n$) 
 in Fig.\ \ref{f:mass}.  For comparison, we also plot threshold stellar masses for galaxies at $z=7-8$ from extrapolations of \cite{Song15}\footnote{Extrapolations for \cite{Song15} were derived from the posterior distribution of their Schechter function parameters, kindly provided by M.\ Song.} and the $z=11.1$ galaxy from \cite{Oesch16}.  Finally, we calculate the expected largest galaxy in the Universe---i.e., using a cumulative number density $n(\ge z)$ such that only one object should exist in the volume of the observable Universe at all redshifts $\ge z$.  These volumes are potentially accessible with all-sky surveys like SPHEREx \citep{Dore14}.  We note that finding a larger galaxy is possible as a result of both sample variance and observational errors (see Appendix \ref{a:code}).

Threshold masses for black holes are shown in Fig.\ \ref{f:mass_bh} and compared to the most massive known quasars and blazars at $z>5$ \citep{Jiang07,Willott10,Volonteri11b,DeRosa11,Mortlock11b,Wu15,Wang15c}.  As bright quasars are detectable in large-area photometric surveys (e.g., the SDSS, \citealt{SDSSQ}, and the CFHQS, \citealt{Willott07}), we calculate mass thresholds at cumulative number densities of $10^{-9.5}$ Mpc$^{-3}$ and $10^{-11}$ Mpc$^{-3}$.  If their virial mass estimates are correct, two quasars, SDSS J1044$-$0125 \citep{Jiang07} and SDSS J010013.02$+$280225.8 \citep{Wu15} likely have $M_\bullet / M_\star$ ratios larger than the $z=0$ relation even if their host galaxies have 100\% $M_\star / M_b$ ratios.  Because blazars have uncertain beaming corrections, we show lower limits assuming $\Gamma=5$ from Fig.\ 5 in \cite{Volonteri11b}. 

\section{Discussion and Conclusions}

\label{s:conclusions}

Several factors limit attempts to rule out $\Lambda$CDM with galaxy or black hole masses, including observational errors and multiple comparisons (see Appendix \ref{a:code}).  For galaxies, there are significant uncertainties in converting luminosity to stellar mass \citep{Conroy09,Behroozi10}; besides systematic offsets, these also induce Eddington/Malmquist bias \citep{Eddington13,Malmquist22} that artificially inflates the number densities of massive galaxies \citep{Behroozi10,Caputi11,Grazian15}. Also present are uncertainties in photo-$z$ codes and priors; improperly chosen, the latter can similarly inflate massive galaxy counts \citep{Stefanon15}.  Just as problematic are multiple peaks in the posterior distribution of $z$, as for the massive galaxy in \cite{Stefanon15}.  We note also the relatively high lensing optical depth at $z>8$, which further boosts the apparent number of massive galaxies \citep{Mason15}.  For a full discussion of other sources of systematic error affecting the stellar mass -- halo mass relation, we refer readers to \cite{Behroozi10}.

Black hole masses are also subject to many uncertainties \citep[see][for a review]{Peterson14}, and virial masses in particular may be overestimates \citep{Shankar16}.  Selecting the largest black holes from a sample with uncertain masses also imposes Eddington bias.  Nonetheless, our limits agree with other approaches that infer large black hole mass --- stellar mass ratios \citep{Targett12,Venemans16}, which are expected due to selecting for luminous, massive black holes \citep{Lauer07,Volonteri11}.  We note in passing that blazars are also subject to the same number density constraints; however, estimates of their number densities are made more complicated due to uncertain beaming factors \citep{Ghisellini09}.

Even so, it is exciting that the highest stellar masses observed in Fig.\ \ref{f:mass} are so close to the limits expected for $\Lambda$CDM.  This suggests that high-redshift galaxy surveys will give lower bounds on the evolution of the halo mass function at $z>8$, which is otherwise very difficult to measure.  Combined with constraints on primordial non-Gaussianities and dark matter from faint galaxies \citep{Habouzit14,Governato15}, \textit{JWST} and \textit{WFIRST} will place very interesting limits on early Universe cosmology.  For SMBHs, Fig.\ \ref{f:mass_bh} provides a simple estimate of whether a given $M_\bullet$ requires an anomalously high $M_\bullet / M_\star$ ratio, potentially bolstering the case for follow-up observations.

Finally, we cite examples of ``unusual'' physics that could be invoked if future observations cross the thresholds outlined here.  We refer to ideas beyond usual prescriptions for supernova feedback and AGN (active galactic nuclei) quenching that limit star formation in current cosmological and zoom-in simulations.  Specifically, we mention positive feedback from AGN (as a precursor to the negative feedback observed via AGN-driven massive gas outflows; \citealt{Gaibler12,Ishibashi12,Silk13,Wagner16}), examples of which are beginning to be found \citep{Zinn13,Cresci15,Salome15}, and a significant duty cycle of hyper-Eddington accretion, increasingly invoked to solve SMBH growth problems at high redshift \citep{Jiang14,Volonteri15,Inayoshi16}.  These processes may be able to increase the ratio of stellar mass or black hole mass to total baryonic mass up to the limit imposed by $\Lambda$CDM (i.e., the cosmic baryon fraction).  Unusual physics that allows accelerated halo growth in overdense regions (e.g., non-gaussianities as in \citealt{Pillepich10}, although standard models are now strongly limited by \citealt{PlanckNonGaussianity}) could also result in overmassive galaxies and black holes that exceed standard $\Lambda$CDM limits.

\section*{Acknowledgements}

We thank Robert Antonucci, Andrea Cimatti, Alister Graham, Ryan Trainor, Marta Volonteri, and the anonymous referee for many extremely helpful comments that greatly improved this paper.  PB was partially supported by program number HST-HF2-51353.001-A, provided by NASA through a Hubble Fellowship grant from the Space Telescope Science Institute, which is operated by the Association of Universities for Research in Astronomy, Incorporated, under NASA contract NAS5-26555.  JS acknowledges support from  ERC Project No. 267117 (DARK) hosted by Pierre et Marie Curie University (UPMC) - Paris 6, France. 

\appendix

\section{Estimating Outlier Probability}
\label{a:code}

\begin{figure}
\vspace{-5ex}
\plotgrace{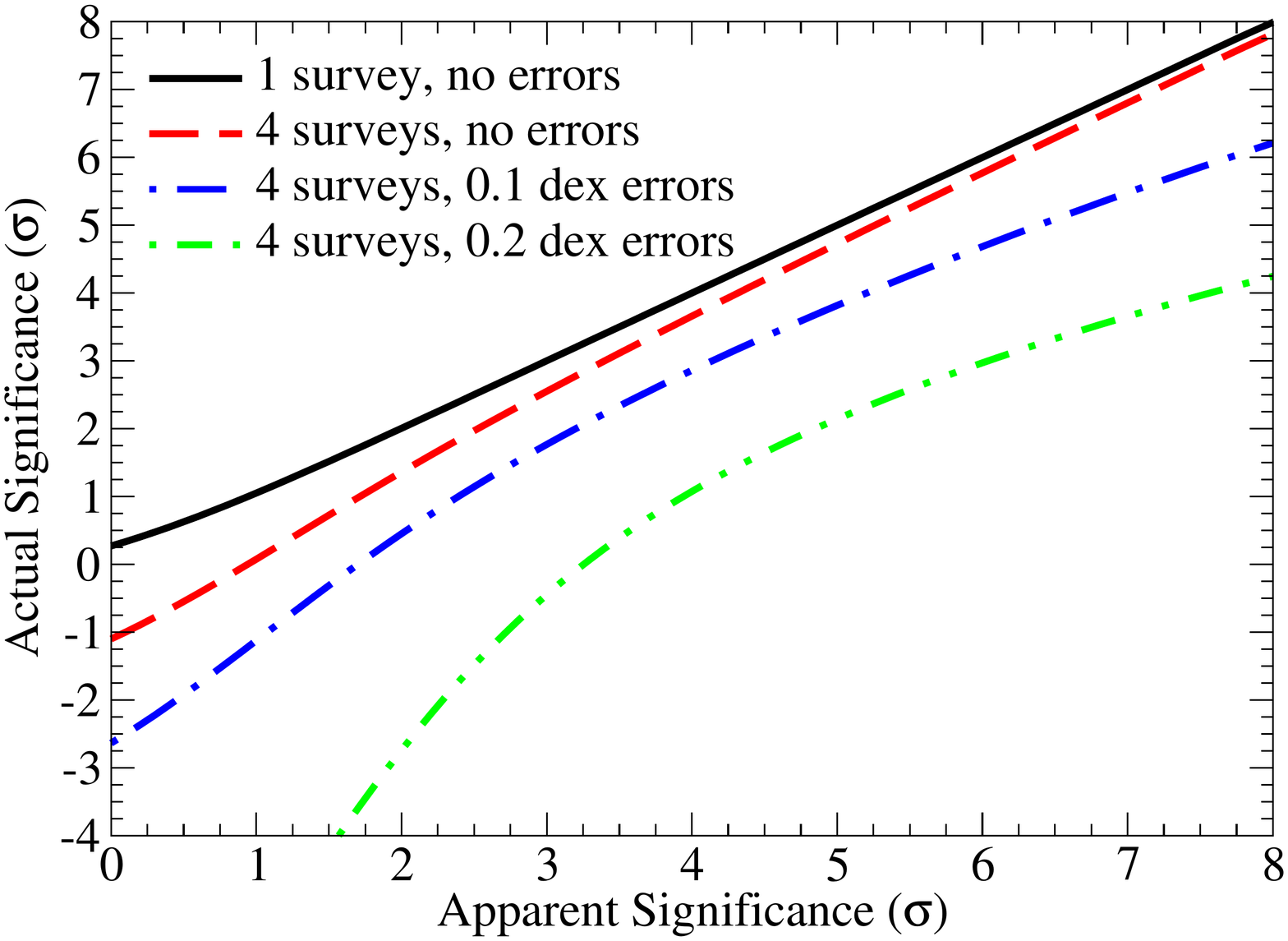}\\[-3ex]
\caption{Even assuming standard $\Lambda$CDM, observational errors and multiple comparisons mean that it is common to find ``overmassive'' haloes.  For a single survey with no errors, the apparent significance (Eq.\ \ref{e:outlier}) corresponds almost exactly to the true significance.  When multiple surveys are conducted, it is common for at least one survey to detect a halo that is slightly overmassive compared to expectations for its volume.  However, even small amounts of observational error can result in significantly disparate true vs.\ apparent significances, due to the steepness of the halo mass function.}
\label{f:true_vs_apparent}
\plotgrace{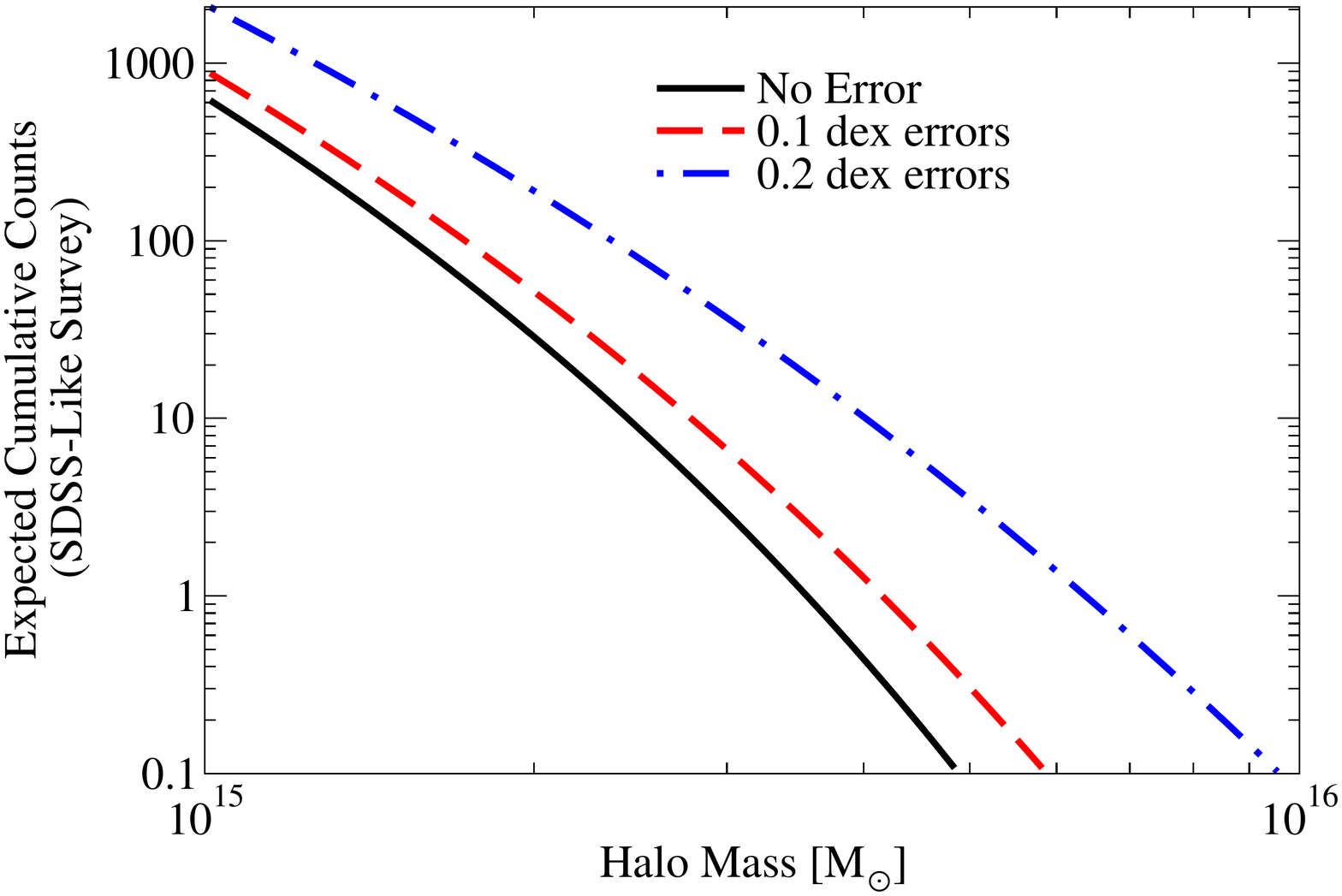}\\[-3ex]
\caption{Effect of log-normal observational errors on total halo number counts in an SDSS-like photometric survey.}
\label{f:obs_errors}
\end{figure}

When a halo is found in a survey with mass $M_h$, it is often labelled ``too massive for $\Lambda$CDM'' if the expected number of haloes is less than some threshold $\epsilon$:
\begin{equation}
V\Phi(m>M_h) \equiv V \int_{M_h}^\infty \phi(m)dm < \epsilon \label{e:outlier}
\end{equation}
where $V$ is the survey volume and $\phi(m)$ is the survey volume-averaged halo number density (per unit mass) for the adopted $\Lambda$CDM cosmology.

The true significance is always weaker than $\epsilon$ would imply; this is due to both observational errors and multiple comparisons.  For extremely rare objects, sample variance from large-scale modes is extremely subdominant to variance from Poisson statistics \citep[see, e.g., the cosmic variance calculator in][]{Trenti08}; hence, we exclude the former effect from our estimate here.  For a single survey, the Poisson chance of observing a halo of mass $M_h$ or larger is
\begin{equation}
P(M>M_h) = 1-\exp\left[-V\int_{M_h}^\infty \left(\int_{0}^\infty \phi(m)P(M|m)dm\right)dM\right] \label{e:survey}
\end{equation}
where $m$ is the true halo mass, $M$ is the observed halo mass estimate, and $P(M|m)$ is the probability density of observing a halo mass $M$ for a true underlying halo mass $m$.

With multiple surveys, the chance increases that one of the surveys will have an ``outlier'' even in a standard $\Lambda$CDM universe.  For the true probability of an $\epsilon$-outlier (according to the definition in Eq.\ \ref{e:outlier}) occurring in at least one of the surveys, we can use the Dunn-\v{S}id\'ak assumption (i.e., multiple fully independent surveys; \citealt{Sidak67}) to estimate:
\begin{equation}
P(\epsilon) = 1 - \prod_{i=1}^n \left[1-P_i\left(M>\Phi_i^{-1}\left(\frac{\epsilon}{V_i}\right)\right)\right] \label{e:survey2}
\end{equation}
where $P_i$ is the equivalent of Eq.\ \ref{e:survey} for the $i$th survey, $V_i$ is the $i$th survey's volume, and $\Phi_i^{-1}$ is the inverse halo cumulative number density for the $i$th survey.  In the regime where $P(\epsilon)<0.05$, this is nearly identical to the Bonferroni limit ($P(\epsilon) \le \sum_i P_i$; \citealt{Dunn58}).

To encourage correcting for these effects, we developed a public code implementing Eqs.\ \ref{e:survey}-\ref{e:survey2}.\footnote{\url{https://bitbucket.org/pbehroozi/lcdm-probability}}  As an example, we compute the relationship between true outlier significance and apparent (Eq.\ \ref{e:outlier}) significance, assuming that an overmassive object were to be found in at least one of the following surveys:

\begin{center}
\begin{tabular}{cccc}
Description & Area (deg$^2$) & Redshifts\\
\hline
SDSS-like & 14555 & $0-0.4$\\
SPT-like & 2500 & $0-1.5$\\
EUCLID-like & 15000 & $0-6$\\
LSST-like & 20000 & $0-3$\\
\end{tabular}
\end{center}

We show results for a range of log-normal observational errors in Fig.\ \ref{f:true_vs_apparent}, presented in terms of more familiar $\sigma$-units.\footnote{Formally, $\sigma \equiv \sqrt{2}\mathrm{erf}^{-1}(2\epsilon -1))$.}  For a single survey with no errors, the apparent significance (Eq.\ \ref{e:outlier}) is close to the true significance (Eq.\ \ref{e:survey}), with the difference arising from the skewness of the Poisson distribution (e.g., an object with average number density of $0.5$ per unit volume will be found in less than half of all such unit volumes, because some volumes will have multiple objects).  Adding multiple surveys as in the table above increases the chance of detecting a massive object in one of the surveys, as expected.  Observational errors result in a very strong effect that is more pronounced for larger surveys.  The effect is analogous to a point-spread function blurring a sharp image; convolving the steep halo mass function with the observational error distribution results in a shallower falloff (e.g., Fig.\ \ref{f:obs_errors}) and therefore an inflated number density of observed massive haloes compared to the underlying true number density (see also Eddington bias; \citealt{Eddington13}).

\newpage

{\footnotesize
\bibliography{master_bib}
}

\end{document}